# Dielectric loss angle for 2D Metal-insulator composite near the percolation threshold

**F.R.Hamou [1], N.Zekri [1], R.Bouamrane [1] and J.P.Clerc [2]**
[1] Laboratoire d'Etude Physique des Matériaux,
Département de Physique, USTO –MB, B.P.1505 El M'naouer, Oran, Algérie.
[2] Ecole Polytechnique Universitaire de Marseille,
Technopôle Château Gombert, 5 Rue E.Fermi, 13453 Marseille – France.
E-mail: nzekri@yahoo.com fr.hamou@gmail.com

**Abstract:**
The frequency dependence of the dielectric loss angle for a metal-insulator composite was shown previously to be an efficient method to experimentally determine the percolation threshold. The statistical properties of this angle are found here to be similar to those of the critical links (supporting the total current) near the percolation threshold close to *dc* and for very high frequencies. A further discussion is provided about the way to connect this angle to the critical links. The second aim of these studies is to connect the frequency dependence of this angle to the proportion of the conductor or dielectric phase in this composite.

**1-Introduction:**
Although percolation theory was extensively investigated within the last four decades, its results, mainly the phase transition and critical exponents, were never experimentally observed. At the beginning of the last decade, Clerc et al. [1,2], showed at the percolation threshold a change in the behavior of a measurable quantity: the dielectric loss angle of metal-insulator composite. It was mainly shown that this angle becomes frequency independent at the percolation threshold of 2D metal-dielectric composites modeled as Resistor-Capacitor networks (absence of a single characteristic time). This behavior suggests a connection between experimentally measurable quantities (such as the loss) and purely fundamentals ones like the number of critical links and the backbone dimension. Indeed, since the infinite cluster is composed of resistors, such links should determine the total resistance behavior for *dc* currents, but *ac* currents at low frequencies also might show a similar behavior for the complex impedance.
Recently, we found two different branches in the distribution of critical links (singly connected links) of a 2D metal dielectric composite; the one is power-law decaying for very small numbers of such links and the other is log-normal [3]. The first branch tends to disappear when the system size becomes large, while the second branch becomes Gaussian as expected by various theoretical investigations at the thermodynamic limit [3]. Since such a distribution was observed for finite size systems, it is interesting to investigate the distribution of a measurable quantity like the loss angle.
This is one of the aims of the present work, where we study numerically for a 2D Resistor-Capacitor network the dielectric loss distribution for very small and very large frequencies at the percolation threshold. The duality between these two limiting frequencies is expected for square networks since at the above threshold we have the same concentration of both components and the insulator becomes conductor for high frequencies. This duality cannot be verified for 3D systems since the percolation threshold (pc about 30%) for low frequencies corresponds to a conducting system for very high frequencies.

Furthermore, the frequency dependence of the loss angle is examined for various metallic fractions. It is mainly shown a concave behavior below the percolation threshold (the system is insulator) with a maximum at the relaxation frequency, while above this threshold the system becomes conducting and the loss behavior becomes convex with a minimum. If we consider this as a peak, its width will diverge at the percolation threshold as the correlation length for *dc* currents. There is then a connection between the geometrical behavior for continuous currents and the frequency behavior for alternating currents. This is the other aim of this work, where we study the behavior of the loss angle band width for different metallic fraction. We try from this investigation to connect the fraction of metallic phase to the dielectric loss band width near the relaxation frequency of the system.

**2- The model and calculation:**

Percolation theory and its application has already been the subject of various works [4]. One of the useful models for the study of the electrical percolation in composites is a resistor random network, where each bond is either metallic or insulator. When the AC electrical response of the composite is investigated, each bond is represented by complex impedance. The metallic bond is modeled by pure resistors describing the electronic mobility and the loss causes by joule effects while the dielectric bond by a capacitor representing the polarization effects between grains. This model is called "RC model" and was developed by Clerc et al [1,12,13], this method has shown to be relevant for understanding the electrical properties of a large variety of inhomogeneous materials, indeed various experiments have been explicitly interpreted by the RC model [7,8,9]. More refined models like the RL-C appeared and devoted to model random metallic-insulator mixtures at high frequencies. In this model each conducting bond of the network is representing by an induction coil with a resistor in series and the insulator by a capacitors. The RL-C model is useful for describing the optical properties of the metal-insulator composites like cermets [11,12,13] who presents an optical transition near the percolation threshold. This model was used to explain and investigate the anomalous absorption observed experimentally for such composites [11,12].In this work we use the RC model to investigate the properties of the dielectric loss angle for 2D metal-insulator composites modeled as square network randomly generated. Each conductor node is modeled as pure resistor $R=10^3$ ohms (real impedance) with a proportion $p$ and the insulator by pure capacitors $C=10^{-8}$ Farad (imaginary impedance) with a proportion $q$:

$$Z_m = R \qquad Z_i = iC\omega \text{ With a characteristic time } \tau = RC$$

The dielectric loss angle is defined by the ratio between the imaginary and the real part of the effective complex impedance of the system:

$$Tg\delta(\omega, p) = \frac{Z''(\omega, p)}{Z'(\omega, p)} \qquad (1)$$

The loss angle of the RC model has the remarkable property near the percolation threshold. Indeed Clerc et al [1,2] have been shown that this angle becomes frequency independent at $p_c$ and assumes the universal value:

$$\delta = \frac{\pi}{2}\frac{s}{s+t} = \frac{\pi}{2}(1-u) \qquad P \approx P_c \quad (2)$$

$$\delta(p,\omega) = \frac{\pi}{2} - \delta(q,\omega) \qquad (3)$$

The critical exponent $u$ is related to the critical exponent $s$ and $t$ by:

$$u = \frac{t}{s+t} \qquad (4)$$

The equality between the exponents s and t implies that ($u$=1/2) and the loss angle ($\delta = \pi / 4$) becomes frequency independent at the percolation threshold. This behavior was observed at the thermodynamic limits only, but remains an open question for finite size systems. This is the aim of the present work

For the calculation of this parameter, a square RC lattice is randomly generated with a concentration $p$ of resistances R. The effective impedance is obtained by using in each mesh a star-triangle transformation (first used by Franck and Lobb [14]). Such a transformation reduces in each step the size of the system up to obtaining a single branch in each direction. Although this method is very fast, the successive divisions appearing can generate in certain situations (where the effective impedance is small) computer errors. In such cases, we compare the result with an exactly solved set of Kirchhoff equations [13,15,16]. For such a computation, we assembled a cluster of 18 nodes (personal computers of 1 GHz speed) in order to reduce computing time. This calculation was performed for different value of $p$ and different size of the system.

**3-Results and discussions:**

**3.1-Statistical behavior of the loss angle:**

This section presents a results obtained by the calculation of the value of Tgδ at the percolation threshold for 10000 samples and for different sizes. The variation of the mean value of Tgδ according to the size of system is represented in figure (1). The behavior of the mean value of Tg δ is in power decaying for sizes lower than 128X128. This is explained by a size effect which means that the value of Tg δ is statistically bad behaved for finite small size system. The second part of the curve corresponds to a saturation of the value of Tgδ who means that for systems of size higher than 128X128 the value of loss angle is statistically well behaved

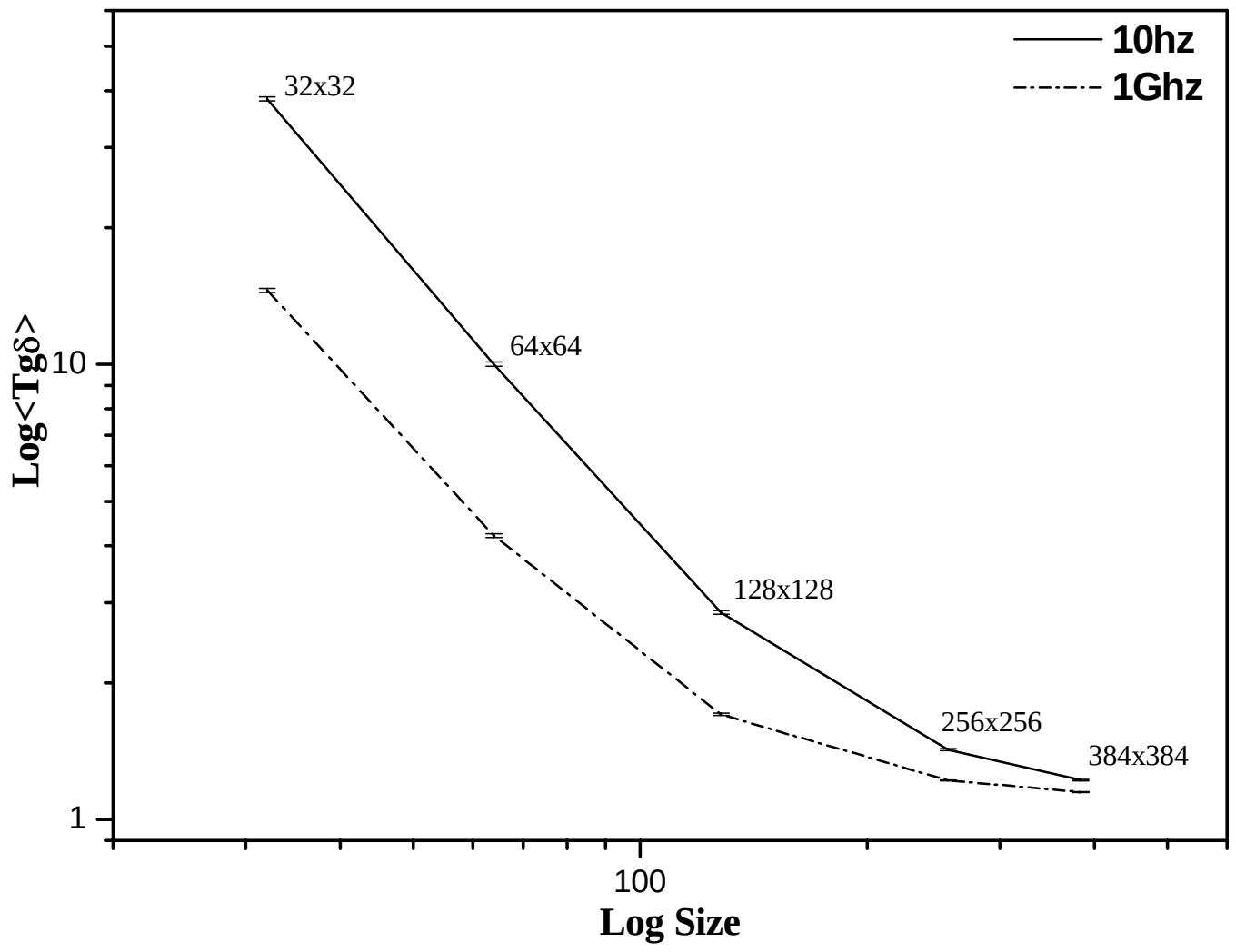

**Figure 1.**Variation of the mean value of Tgδ according to the size of system at the percolation threshold.

The distribution of Tg(δ) is studied here in the asymptotic limits of very high and very small frequencies. It is worth noticing that while for small frequencies (nearly *dc* case) the capacitance corresponds to the insulating component, for high frequencies it becomes highly conducting in comparison to the pure resistance. Since the percolation threshold for two dimensional square lattices is 50%, we expect the properties of the infinite conducting (R) and insulating (C) clusters identical. This allows us to expect the same statistical behaviors in the two frequency limits.
The distributions of Tgδ, obtained from 10000 samples realizations at *Pc*, are shown in figure (2) at 10Hz and figure (3) at 1GHz for various sizes. These figures confirm the similarity between the two frequency limits. The distribution presents two branches. The first, close to zero which decrease in power who corresponds to configurations with very small loss angles and another in Log normal. The first branch tends to disappear when the size of the system becomes increasingly large and becomes Gaussian near a value equal to 1 ($\delta = \pi / 4$ value predicted for an infinite system [3]) for large sizes higher than 128. This explains the fact that the value of Tg δ stabilizes around a more probable value for systems of very large sizes. In this case the mean field theory is applicable for systems of significant sizes contrary for the small sizes.

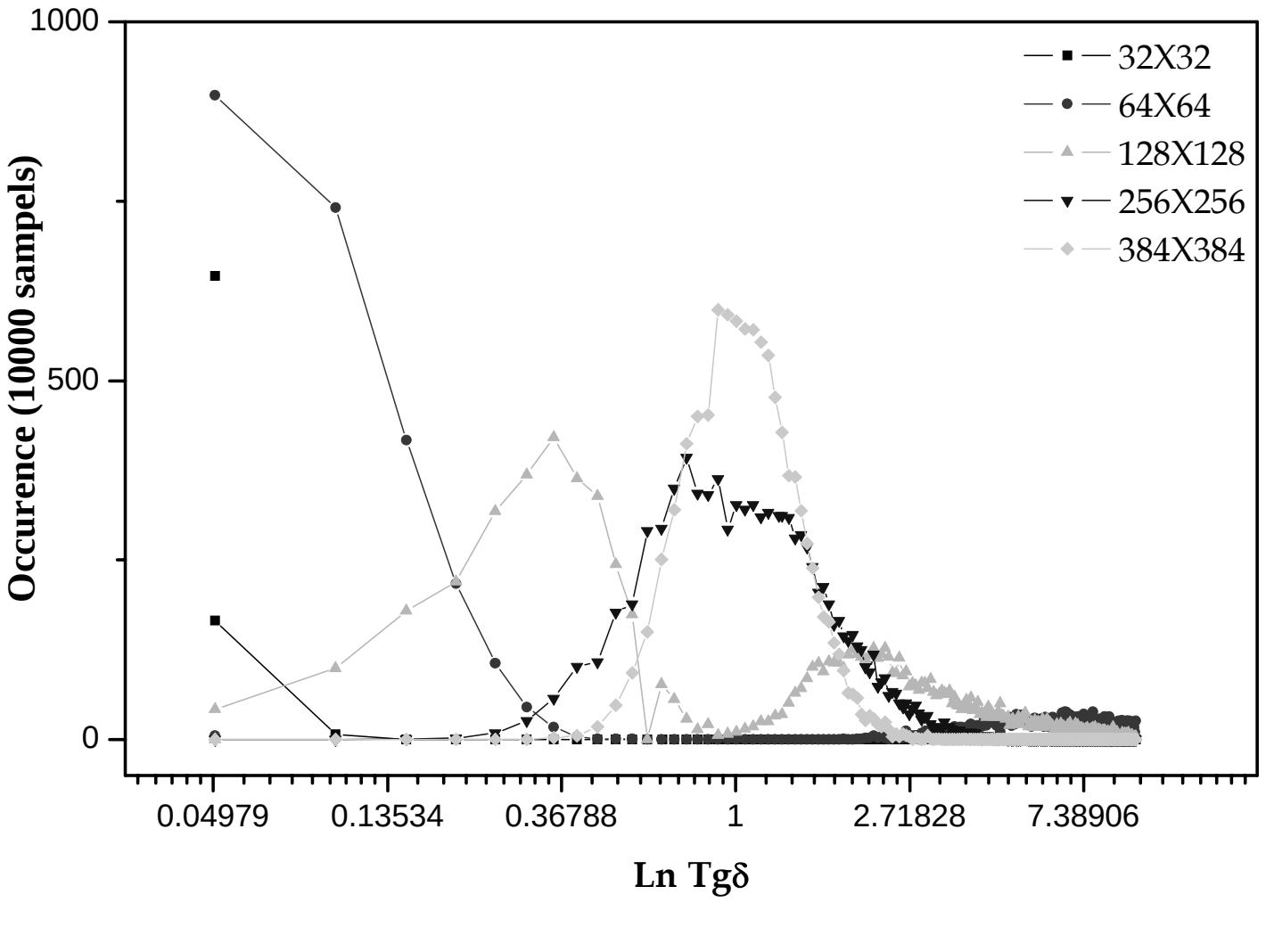

**Figure 2.** Occurrence of Tg δ at 10Hz and $p=p_c$

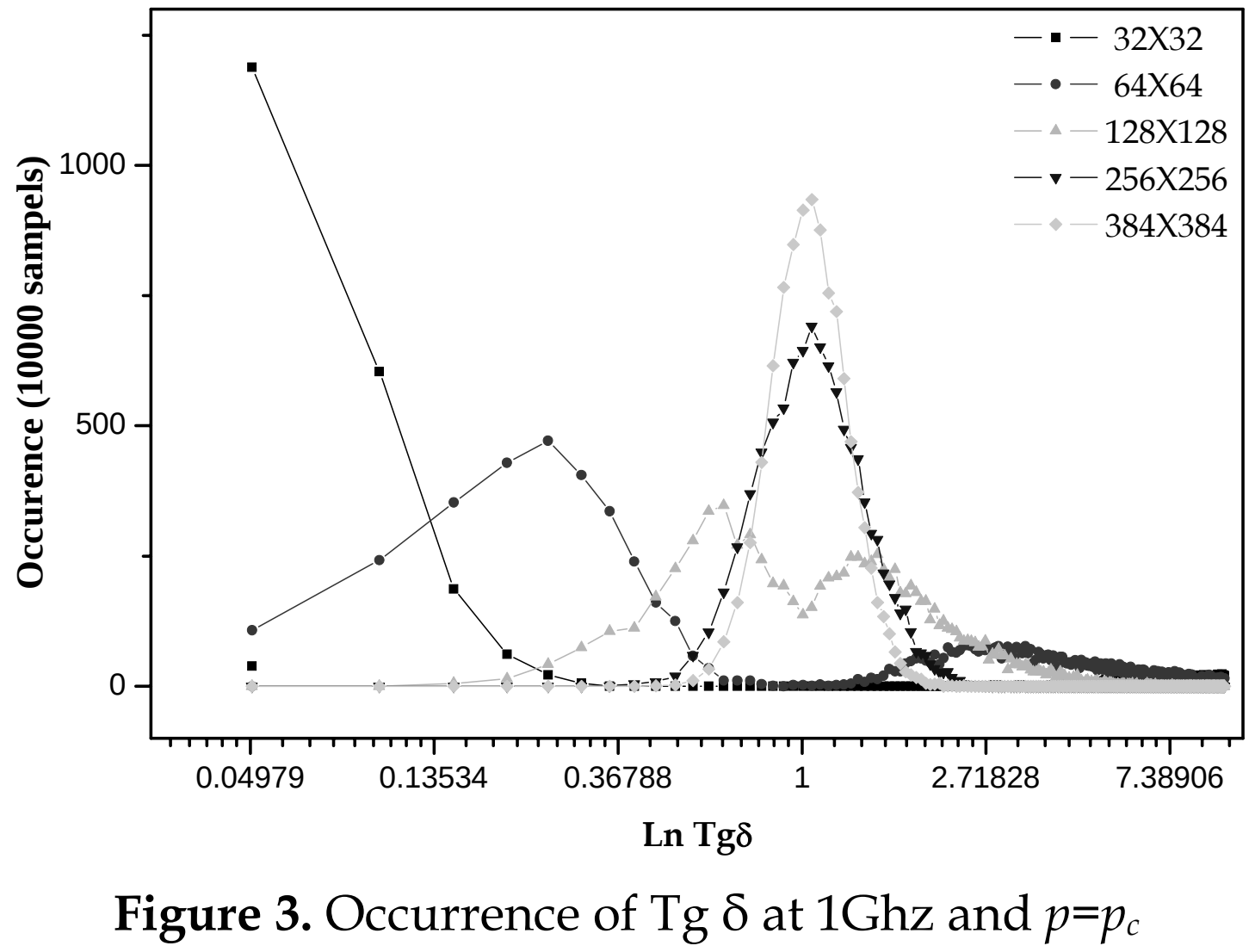

**Figure 3.** Occurrence of Tg δ at 1Ghz and $p=p_c$

The same phenomenon is noticed for the statistical behavior of the critical bonds recently observed by Clerc and Zekri [3]. Indeed, they noted that for finite size systems, the distribution of the mean number of critical bonds (those carrying the totality of the current) presented two branches: One is power-law decaying for samples corresponding to very small numbers of red bonds, and the other one is Poisson-like for the same size. The first branch tends to disappear (or to amalgamate with the other) and becomes Gaussian when the size increases.

They explained this behavior by very probable configurations where the percolate cluster does not contain this type of bonds, whereas these configurations are rare when the size of the system tends towards the thermodynamic limit. This similarity encourages us to think that the loss angle Tgδ which is a measurable size is directly connected to the number of critical bonds (fundamental quantity of the percolation theory). It is thus obvious that this loss angle accounts the structure of the infinite cluster (backbone) in the system.

This consolidates the predictions in frequency of Clerc et al [1,2] who show that the spectrum of the loss angle changes concavity with the transition from percolation and becomes straightforwardly independent of the frequency exactly at the percolation threshold. We will examine in detail this effect in the next section.

**3.2- The properties of the loss angle near the percolation threshold:**

The previous section has been concerned with a statistical property of the dielectric loss angle for a 2D square metal-insulator composite. We shall describe in this section some properties of this angle for such systems, modeled as a bond percolation lattice. Many theoretical and experimental studies were devoted for these properties [7,8,9,10] were many experiments have been interpreted in terms of the RC model in a percolating geometry. In this section we calculate the frequency dependence of this angle to the proportion of the dielectrics bonds. Indeed, our calculation was performed for different proportion of dielectric phases and different size of 2D-RC systems. Figure 4 present the numerical values of the loss tangent against the frequency for different proportion of dielectric bonds. This result has been predicted by Clerc et al [1,2] by the effective medium approximation or by Laugier and Luck [1] using the transfer matrix. In particular, as a consequence of equation (3), the plot of figure 4 is symmetric whereas all curves, both above and below the percolation threshold present a peak near the relaxation frequency and are very close to each other at higher and lower frequencies. The curve that corresponds to the percolation threshold is frequency independent.

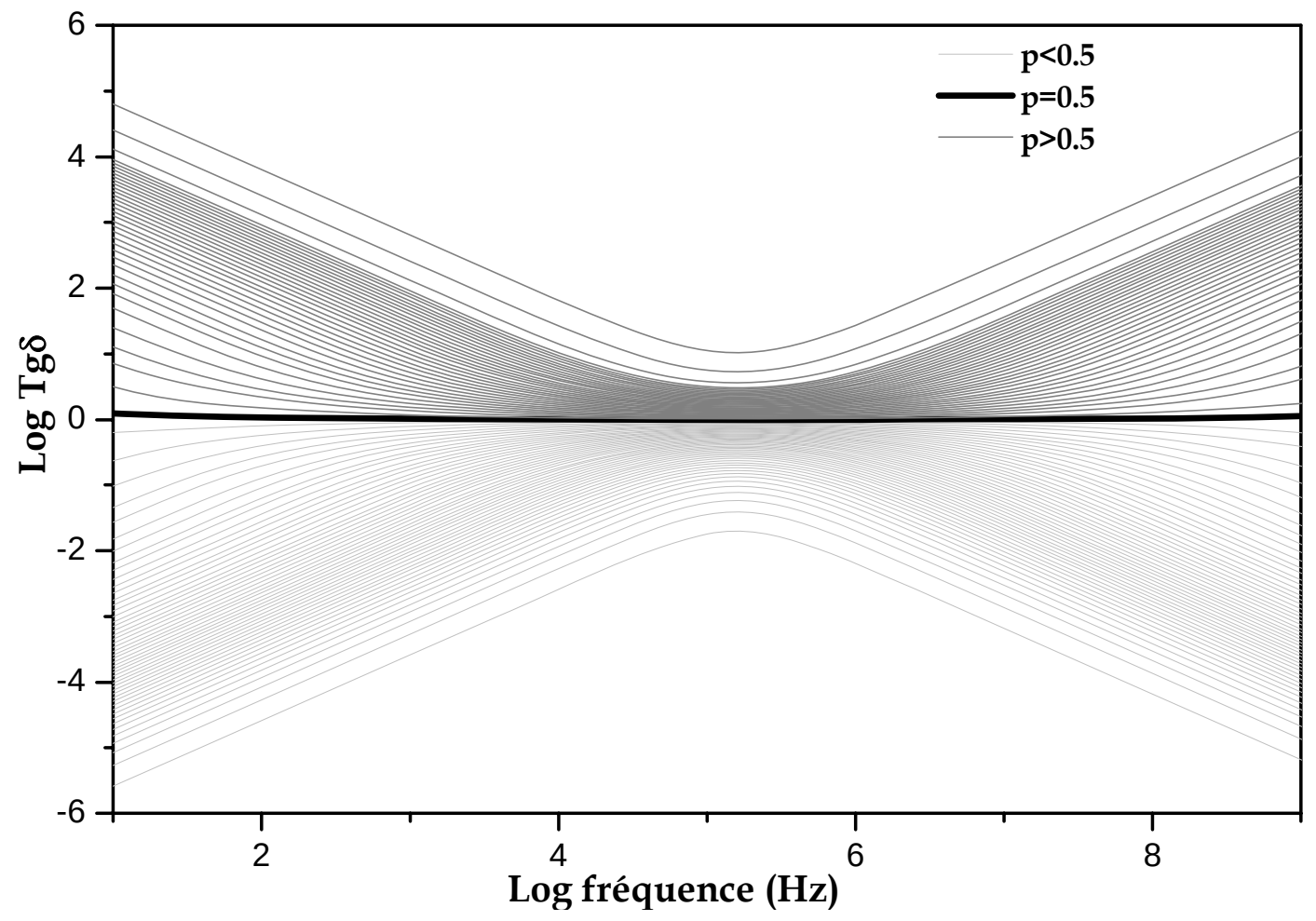

**Figure 4.** The loss tangent versus log frequency for different value of *p* for 384x384 systems.

The analysis of the behavior of the loss angle δ for each proportion $p$ enabled us to note that the full width at the half of the maximum was related to the proportion of the dielectric phase. The existence of the symmetrical behavior between the insulating and conducting state in material is very noticed in figure 5. It is noted that for reverses proportions (R:10%, C:90%) and (R:90%, C:10%), the width at the half of the maximum is practically the same one.

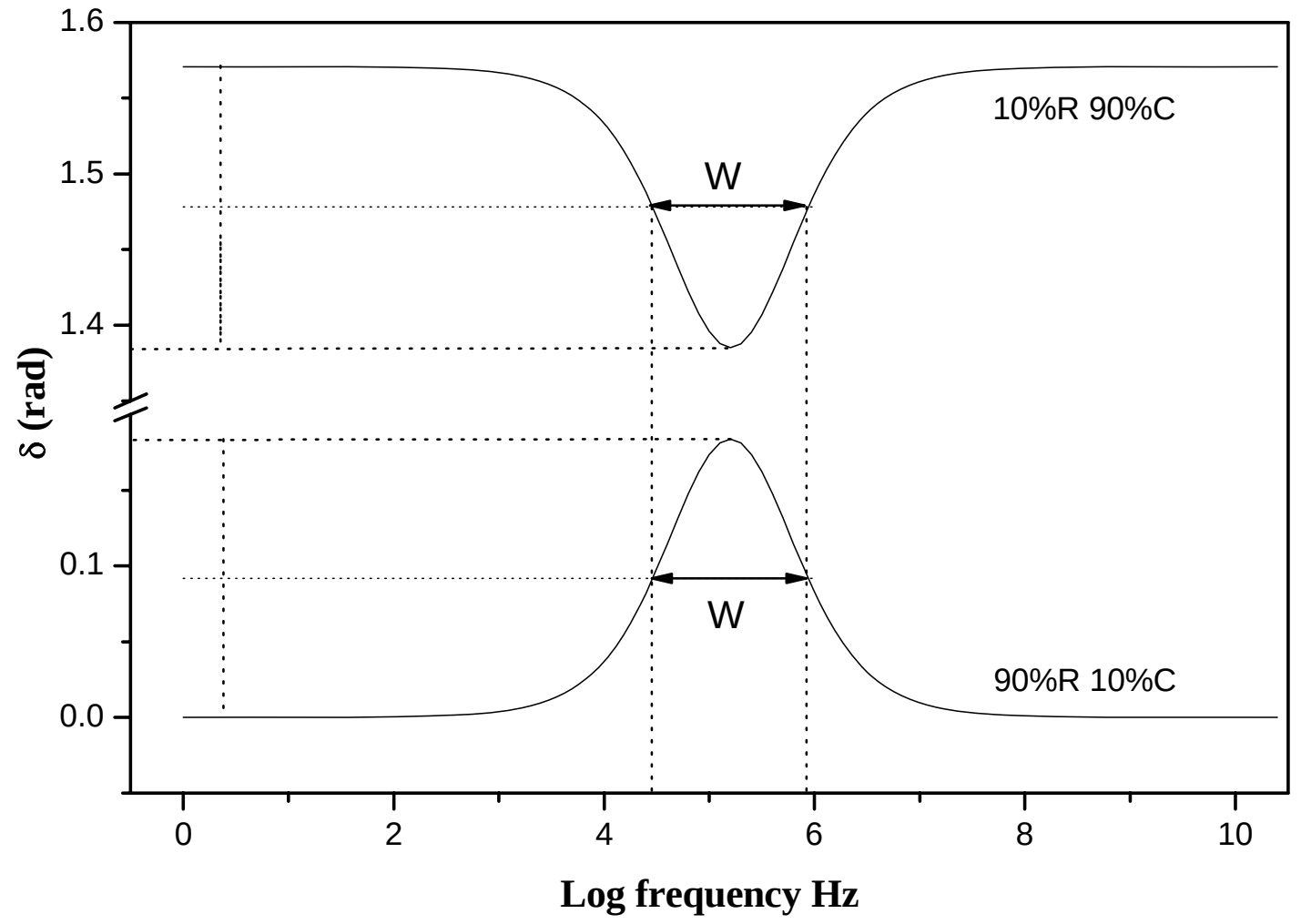

**Figure 5.** The loss angle δ versus log frequency for reverses proportions.

We calculated this width for all the proportions $p$ of the insulating grains, this variation is represented in figure 5 for different systems size. The full width at the half of the maximum represents the difference between the times periods (inverse of frequencies) who delimited the width. The asymptotic behavior is clearly observed in figure 6 for different sizes of system, the width W diverge as the percolation threshold is approached from both sides.

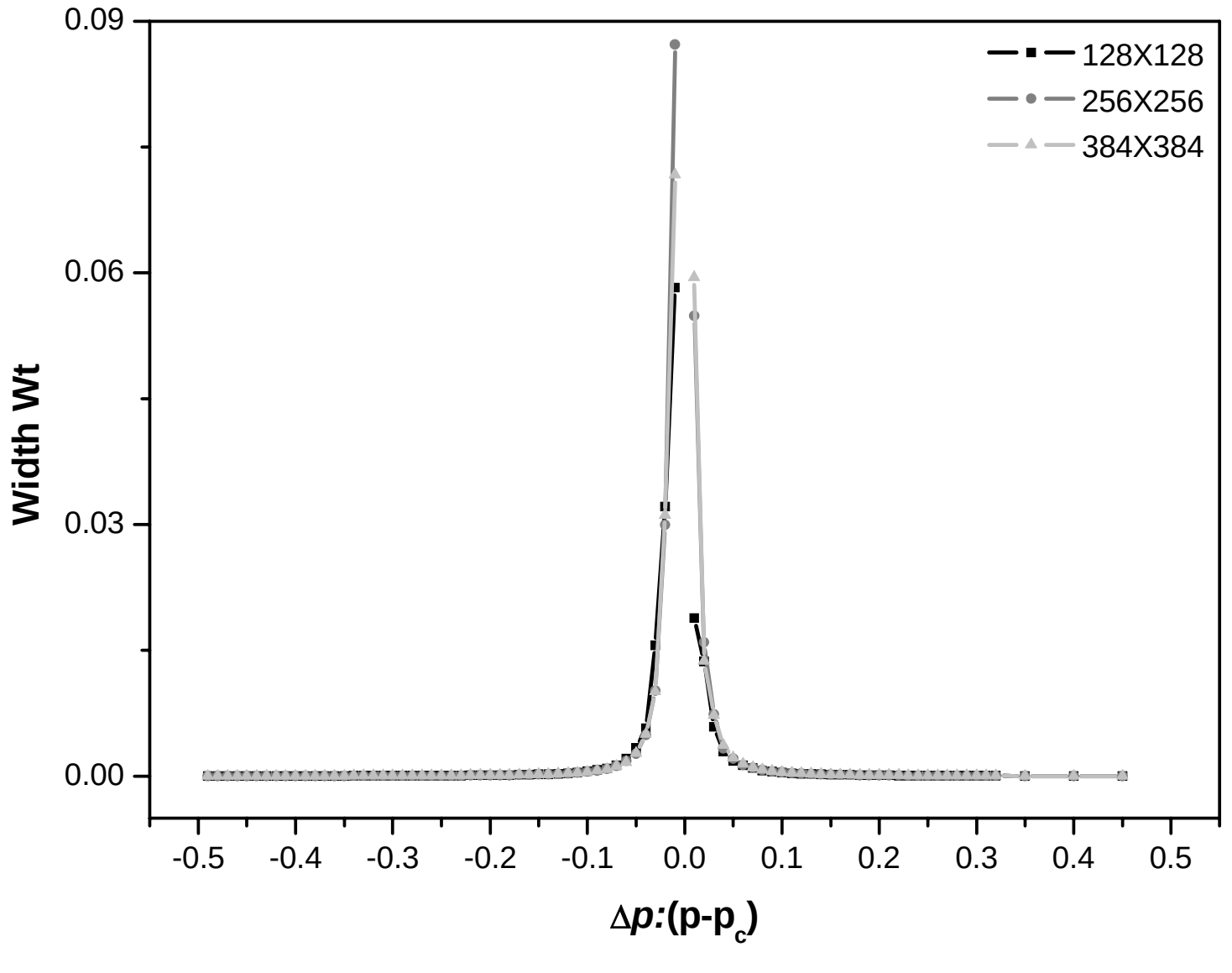

**Figure 6.** Variation of the WHM versus $\Delta p$ for different sizes of system.

This behavior has been observed for different parameter in percolating geometry, one of this parameter is a critical time scale defined by [2]:

$$\tau^* \sim \omega_0^{-1} \left|\Delta p\right|^{-(s+t)} \qquad (5)$$

Indeed, the theory of critical dynamics predicts the existence of a divergent time scales

$$\tau^* \sim \xi^z \qquad (6)$$

Where $\xi$ is the static correlation length and $z$ is the dynamical critical exponent defined by:

$$z = \frac{s+t}{\nu} \qquad (7)$$

$t$ is a dynamical exponent whereas $s$ is a static one, and are believed to be universal and independent of each other. In two dimensions a duality argument established that $s$ and $t$ are equal [17]. The critical exponent can be calculated by means of the finite size scaling [18]. Variants of this method have been used by Lobb and Frank [19] and other [20,21] in percolation conductivity in two dimensions

The variations of the width (WHM) are fitted by the following expression (8) that resembles to the relation (5):

$$W = \frac{1}{\omega_0 \left| p - p_c \right|^{2x}} \qquad (8)$$

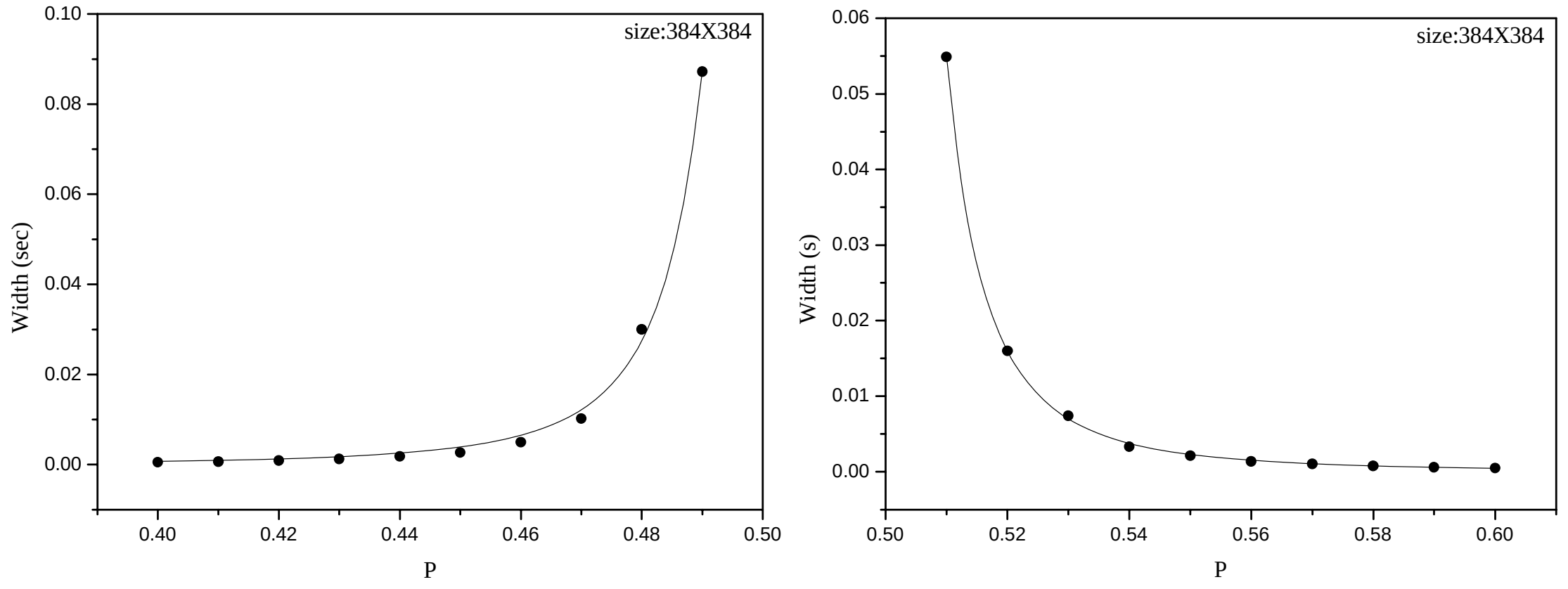

(a) (b)

Figure 7. The fit plot of the width versus the proportion *p*.

$(a)\ p < p_c \quad \omega_0 = 10^5 rad.s^{-1} \quad p_c = 0.50919 \pm 0.00191 \quad x = 1.38042 \pm 0.0264$

$(b)\ p > p_c \quad \omega_0 = 10^5 rad.s^{-1} \quad p_c = 0.49415 \pm 0.00016 \quad x = 1.26035 \pm 0.00408$

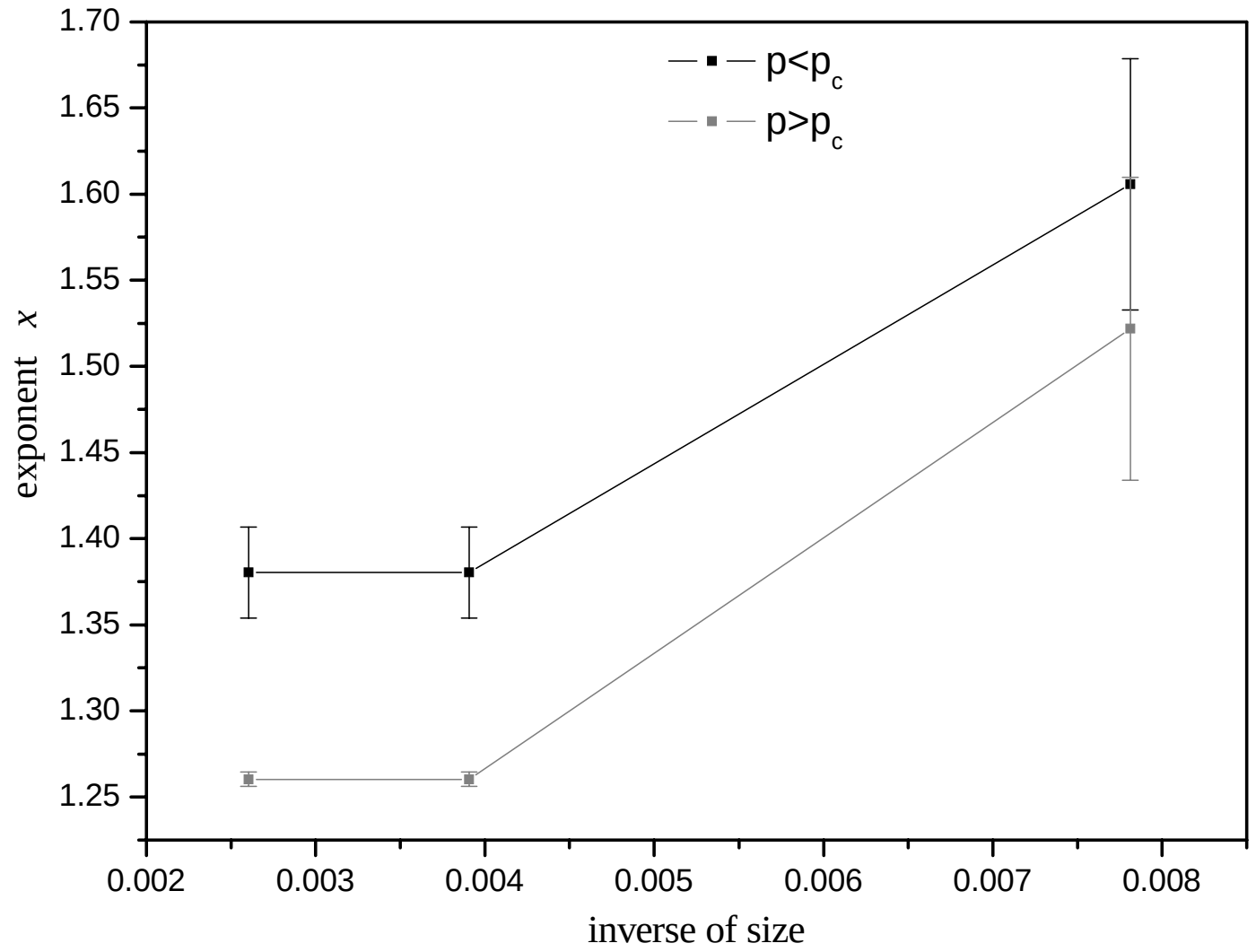

Figure 8. Variation of the exponent x versus the inverse of the size.

The extracted exponent from the fit for a 384X384 system was very close to the dynamical exponent *t* defined in different works [17,18,19,20,21]. More system sizes were investigated by fitting their points and the variation of the exponent was plotted in figure 8 versus the inverse of size system. This figure shows that the value of the exponent tends to stabilize near the predicted value of *t*. This result enabled us to suggest that the width at the half of the maximum of the loss angle behaved like the critical time defined in theory of critical dynamics.

$$W \sim \omega_0^{-1} \left|\Delta p\right|^{-(s+t)} \qquad (8)$$

Theses results permit us to correlate this width (WHM) to the static correlation length. Indeed, the correlation length was related to the critical time by the relation (6).

$$\begin{aligned} W \sim \tau^* \sim \omega_0^{-1} \left|\Delta p\right|^{-(s+t)} \\ W \sim \tau^* \sim \xi^z \end{aligned} \qquad (9)$$

In the previous section, we have concluded that the loss angle which is a measurable size is directly connected to the number of critical bonds and accounts the structure of the infinite cluster (backbone) in the system. This is another evidence to consolidate that the loss angle was directly correlated to this angle, since that the loss angle reflects the proportion of the resistive bond (who constitute the cluster) in the materials through the width at the half of the maximum of the loss angle.

**3.3- Conclusion:**

From the above discussion it is evident that we have largely explained our data with the RC model. It was possible to relate a measurable quantity like the loss angle to the theoretical size like the static correlation length. This was established by a key scaling law with the two static and dynamical exponents s and t who tends to the predict values. Which wants to say that will be can a means which can be applied for the direct measurement of the number of critical bonds. This enables us to say that it possible from an experimental point of view to estimate the metallic or dielectric proportion through the measurement of the width of the peak of the loss angle near the relaxation frequency. Indeed, it has been shown [22] recently how insulating particles with nano scale metal coatings enable dielectric relaxation to a lower frequency regime. It is demonstrated that a model in which the metal coating is assumed to be near the 2D percolation threshold can provide an improved validation to experimental results and the relative peak bandwidth near the relaxation frequency decreases slowly with increasing coating thickness.

4.4- References: